\documentclass[a4paper,11pt,twoside]{article}
\usepackage[INRIA]{lipinria}

\usepackage{pgf}

\newcommand{\matrice}[2]{\left[ \begin{array}{c} #1 \\ #2 \end{array} \right]}
\newcommand{\quatrice}[4]{\left[ \begin{array}{cc}
                                     #1 & #2 \\ #3 & #4 \end{array} \right]}
\newcommand{\barre}{\;|\;}
\newcommand{\Select}[2]{\{#1 \barre #2\} }
\def\bbbn{\mathrm{I\!N}}
\newenvironment{proof}{\begin{quotation} {\bf Proof}}
{\noindent\rule{0.5em}{2.0ex}\end{quotation}}

\begin{document}

\RRInumber{6193}

\RRItitle{Elementary transformation analysis for Array-OL}
\RRItitre{Analyse de transformations élémentaires pour Array-OL}

\RRIdate{May 2007}

\RRIauthor{Paul Feautrier}

\RRIthead{Analysis for Array-OL}
\RRIahead{P. Feautrier}

\RRIkeywords{Array-OL, multidimensional signal processing, program analysis}
\RRImotscles{Array-OL, traitement du signal multidimensionnel, analyse de 
programme}

\RRIabstract{Array-OL is a high-level specification language dedicated to
the definition of intensive signal processing applications. Several tools
exist for implementing an Array-OL specification as a data parallel program.
While Array-OL can be used directly, it is often convenient to be able
to deduce part of the specification from a sequential version of the 
application. This paper proposes such an analysis and examines its
feasibility and its limits.
}

\RRIresume{Array-OL est un système de spécification de haut niveau spécialisé
dans la définition d'application de traitement du signal intensif. Il existe
plusieurs ateliers qui transforment une spécification Array-OL en un programme
à parallélisme de données. Bien que Array-OL puisse être utlisé tel quel,
il est souvent intéressant de pouvoir déduire ses paramètres d'une version
séquentielle de l'application. Ce rapport propose une telle analyse et en
examine la faisabilité et les limites.
}

\RRItheme{\THCom}
\RRIprojet{Compsys}

\RRImaketitle
\bibliographystyle{plain}

\section{Introduction}

In the Array-OL formalism \cite{Boul:07,Eric:07}, 
a program is a network of processes which
communicate through shared arrays. A process is made of one or more
parallel loops. At each iteration of these loops, a task (or elementary
transform) is executed. The elementary transform may contain one or more
loops, which are executed sequentially.

The execution of an elementary task can be decomposed into three steps:
\begin{itemize}
 \item Move portions of the input array(s) (regions) 
to the local memory of the
processor executing the task.
 \item Execute the elementary transform and generate portions of the output
array(s).
 \item Move the results to the output array(s).
\end{itemize}

In order to simplify code generation, the input and output regions must
move uniformly across the shared arrays. It is
admissible that each elementary transform use only a subset of regularly
spaced entries in the input and output regions. In the present version
of the software, regions must not overlap, as this would precludes
parallel execution of the outer loops. The useful elements of a region
are collected in a pattern, which must be a rectangular parallelepiped
of fixed size.

The Array-OL formalism may be used directly. The programmer is responsible
for constructing the elementary transform, identifying the input
and output regions, checking parallelism and specifying the regions
parameters. Another possibility is to infer the Array-OL specification
from a sequential version of the program. This requires the solution of 
three problems:
\begin{itemize}
 \item Rewriting the sequential program in such a way that the outer loops
have no dependences.
 \item Deducing the shape and size of the regions from an analysis of the
array subscript functions.
 \item Rewriting the sequential code by substituting pattern accesses to
the original array accesses.
\end{itemize}

This note is dedicated to a proposal for the solution of the second 
and third problems.
The assumption is that one is given the sequential code, together with
a list of input and output arrays, and an indication of which loop(s)
are to be considered as the outer (repetition) loop(s).

\section{Paving}

Let $A$ be an input or output array and let its occurences
in the sequential code be numbered from 1 to $N$. 
Let $r$ be the counter(s) of the
repetition loop(s), and let $j^k$ be the counter(s) of the inner loop(s)
that surround occurence $k$ of $A$. Let $e^k(r,j^k)$ be its subscript
function. $e^k$ is a vector function whose dimension is the rank of $A$.

To be amenable to an Array-OL implementation, the subscript function
$e^k$ must be affine in $r$ and $j^k$. A convenient way of checking 
this property consists in computing the two Jacobian matrices:

\[ P^k = (\frac{\partial e^k_\alpha}{\partial r_\beta}) \;\;\;
   B^k = (\frac{\partial e^k_\alpha}{\partial j^k_\beta}), \]
checking that they do not depend on $r$ or $j^k$, and verifying the
identity:
\[ e^k(r, j^k) = P^k r + B^k j^k + e^k(0,0) .\]

In Array-OL terminology, $P^k$ is the paving matrix, and $e^k(0,0)$ is
the origin of the paving. The elements of these entities may be numbers,
or they may depend on constants, which must be given numerical values
just before code generation. References with different paving matrices 
may be separated by arbitrary distance in the source or target array;
it is not possible to group them efficiently; they must be implemented
as separate channels.

\begin{quote} \small
In the following example:

\begin{verbatim}
myTE( in[][], out[]){
for(i=0;i<7; i++)  // boucle TE
{
for ( k=0;k<11;k++)
{
S=0;
    for(j=0;j<100;j++)
    {
    S+=  in[0][j+11] * in[i+1][k+j];
    }
    out[ i][k]=S;
}
}
\end{verbatim}
there are two references to {\tt in} with repective subscript functions
$e^1(i,k,j) = \left(\begin{array}{c} 0 \\ j+11 \end{array} \right)$
and $e^2(i,k,j) = \left(\begin{array}{c} i+1 \\ k+j \end{array} \right)$.
The corresponding paving matrices are 
$P^1 =\left(\begin{array}{c} 0 \\ 0 \end{array} \right) $ and
$P^2 =\left(\begin{array}{c} 1 \\ 0 \end{array} \right) $. Hence, the two
accesses must be handled separately.
\end{quote}

In the following, I assume that accesses to $A$ have been partitioned
according to their paving matrix, and consider only one partition at a time.
The size of the repetition space is deduced simply from the bound(s) of
the elementary transform
 loop(s). In the Spear/DE implementation of Array-OL, there may be
further constraints on the paving matrix (e.g. that it be a permutation
of a diagonal matrix).

\section{Pattern and fitting}

A pattern is a compact specification of all the elements of an array
that are accessed, with references having the same paving matrix,
in one iteration of the external loop(s).

\begin{figure}
\begin{center}
\pgfimage[width=0.8\textwidth]{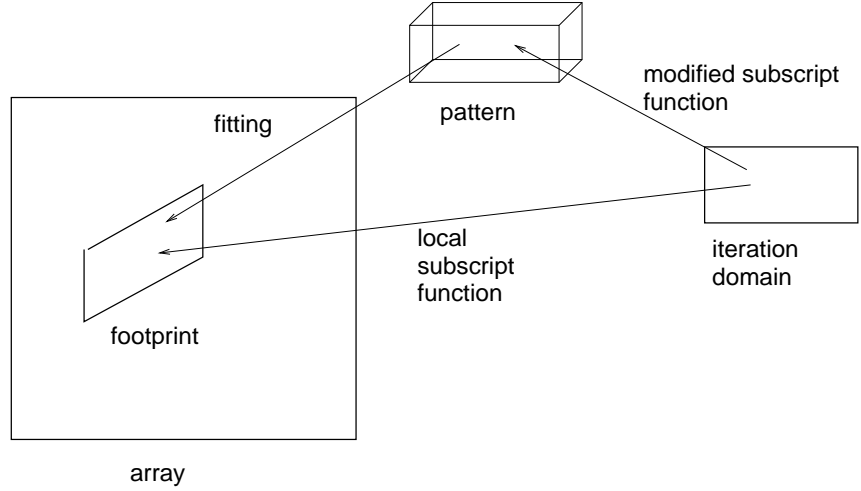}
\end{center}
\caption{Data access in Array-OL} \label{areole}
\end{figure}

When discussiong patterns, one has to consider three frames of reference
(see Fig. \ref{areole}).
The first one is the original (input or output) array. Its dimension is the
rank of the array, noted $|A|$, 
and its coordinates are called \emph{subscripts}. 
The shape of an array is always a (hyper-) rectangle.

The
second frame of reference is the iteration space of the inner loops of
the elementary transform. Its dimension is the number of loops enclosing
the reference, noted $d^k$, 
and its coordinates are called \emph{loop counters}. There
may be as many iteration domains as there are references, or several
references may share the same iteration domain. The shape of an iteration
domain is arbitrary. The only requirement in the present context is to be
able to construct its vertices, either because the iteration domain is
rectangular, or because it can be expressed as a convex polyhedron with
parameters in the constant terms only. 
The iteration domain of reference $k$ will be denoted
as $D^k$ in what follows.

The third frame of reference is the pattern. According to Boulet \cite{Boul:07}
the pattern is always of rectangular shape. The pattern associated to
reference $k$ is denoted by $T^k$ and its dimension is $p^k$.
The associated fitting matrix, $F^k$, connects the pattern space to the
array space and its dimension, accordingly, is $|A| \times p^k$.

The relation of these objects are as follows. Firstly, the local subscript
function $f^k(j^k) = B^k j^k + e^k(0,0)= e^k(0, j^k)$ 
gives the coordinates
of an array cell relative to the reference point $P^k.r$ which
moves according to the paving matrix.

Next, the image $f^k(D^k)$ is the \emph{footprint} of reference $k$.
Its shape is arbitrary. The images of the vertices of $D^k$ by $f^k$
form a superset of the vertices of the footprint; a representation as a 
convex polyhedron can be recovered by one application of the
Chernikova algorithm \cite{Schr:86}. 

Lastly, the image of the pattern by the fitting matrix must enclose the
footprint, and it must be feasible to retrieve a datum from the pattern instead
of the original array. This implies that there exists a function 
$\phi^k$ from $D^k$ to $T^k$ such that for every iteration vector 
$j^k \in D^k$, $f^k(j^k) = F^k \phi^k(j^k)$. In the text of the elementary
transform, $\phi^k$ must be substituted to $e^k$ in reference $k$ to $A$.

As one may see from this discussion, while the iteration domain and footprint
are fixed once the sequential program is given, the choice of the pattern
and fitting matrix are somewhat arbitrary. There are two obvious solutions:
in the first one, the pattern is the smallest rectangular box
enclosing the footprint, the fitting matrix is the identity, and the
subscript function is not changed.
In the second solution, the pattern is isomorphic to the iteration domain
(provided it is a parallelepiped), $B^k$ is the fitting matrix, and the
new subscript function is the identity. 

In signal processing applications, it is often the case that several references
to the same array have similar subscript functions; constructing only
one pattern for several references is an interesting optimization.
However, this should not be obtained at the cost of a large overhead
in the size of the pattern. In other word, the number of useless elements
in the pattern must be minimized. Useless elements come from two sources:
\begin{itemize} 
\item A subscript matrix which is not of full row rank: the
pattern will have more dimensions than the footprint.
\item A subscript matrix whose determinant is not of modulus one: there will
be holes (unused elements) in the footprint. The inverse of the determinant
gives an asymptotic evaluation of the ratio of useful elements.
\end{itemize} 

The next section presents a method for computing a pattern and a fitting
matrix in the general case (many references). This method can only be
applied if all elements of the matrices $B^k$ and the vectors $b^k$ have
known numerical values. Section \ref{approx} presents fail-soft solutions
for cases in which these elements depend on unknown parameters.

\section{The General Case}

The basic observation is that a conservative estimate of the footprint
can be obtained by computing the projection of each iteration domain
by the associated subscript function, then constructing a convenient
superset of the union of these projections. One practical method
consists in projecting the vertices of the iteration domains. One then gathers
all such projections, and constructs their convex hull by familiar
(e.g., Chernikova's) algorithms.

To reduce the size overhead, one should notice that a useful point 
for reference $k$ also
belongs to the lattice which is generated by the column vectors of $B^k$.
Hence, $B^k$, properly simplified (see later) could be used as the fitting
matrix. However, in the case of several references, we have to combine
several lattices into one, since each pattern has only one fitting
matrix. As an illustration of this construction, consider the one dimensional
case. A one-dimensional 
lattice is simply a set of regularly spaced points. Combining
two lattices generates a lattice whose spacing is the gcd of the component
spacings. The many-dimensional equivalent of the gcd is the construction
of the Hermite normal form of the subscript matrices.

Let $\Lambda(B,b)$ be the lattice generated by $B$ with origin $b$, i.e.
the set of points $\Select{Bx+b}{x \in \bbbn^d}$. Let $L^1 = \Lambda(B^1,b^1)$
and $L^2 = \Lambda(B^2,b^2)$ be two such lattices. I claim that the union of
$L^1$ and $L^2$ is included in the lattice 
$L = \Lambda([B^1 B^2 (b^2 - b^1)], b^1)$.
\begin{proof}
Let $B^1 . x + b^1$ be a point of $L^1$. We have:
\[ B^1 . x + b^1 = B^1 . x + B^2 . 0 + (b^2 - b^1) . 0 + b^1 \]
hence $B^1 . x + b^1$ is in $L$. Similarly:
\[ B^2 . y + b^2 = B^1 . 0 + B^2 . y + (b^2 - b^1) . 1 + b^1 .\]

I conjecture that $L$ is the smallest lattice which includes $L^1$ and $L^2$.
The proof is obvious if the $b$s are null. The general case is left for
future work.
\end{proof}

The construction can be extended to any number of component lattices.
The resulting matrix is $[B^1 \ldots B^N (b^2 - b^1) \ldots (b^N - b^1)]$
and the origin is $b^1$. Furthermore, $b^1$ can be moved to the origin
of the paving and hence taken as 0 when computing the fitting.

In case where $B$ has been obtained by mixing many references, it must be 
simplified before being used for an Array-OL specification.

The starting point of this simplification
 is the \emph{row echelon} form of $B$.
One can show (see the appendix) that there exists two unitary matrices
$P$ and $U$ such that:
\[ B = P \quatrice{H}{0}{C}{0} U ,\]
where $H$ is a square upper triangular matrix of size $r \times r$ with
positive diagonal coefficients, $C$ is arbitrary, and both 0 represent null
matrices of appropriate sizes. $r$ is the row rank of $B$.
Furthermore, $U$ can be partitioned, row wise, in two matrices of size
$r \times d$ and $(d-r) \times d$, $U = \matrice{U'}{U''}$.

Let $j$ be a point in the iteration domain of the inner loops. The 
corresponding point in the footprint is:
\begin{eqnarray}
 Bj & = & P \quatrice{H}{0}{C}{0} \matrice{U'}{U''} j\\
    & = & P \matrice{H}{C} (U' j)
\end{eqnarray}
One possible interpretation of this formula is that the pattern for the 
current reference is the image of its iteration domain by $U'$, and that
the corresponding paving matrix is $P \matrice{H}{C}$. 
In the body of the elementary
transform, accesses to $Bj$ in the input or output array have to be 
replaced by accesses to $U'j$ in the pattern. It may be that the pattern 
computed in this way is not rectangular, in which case it must be ``boxed'' 
by computing the component-wise minima and maxima of its extreme points.
The dimension of the pattern is $r$.

It is interesting to notice that this general solution reduces to one of the
approximate methods above in special cases. If $B$ is unitary,
then its row echelon form is the unit matrix. In that case, the pattern is 
the footprint, eventually extended to a rectangular box and the fitting matrix
is the identity. Conversely, if $B$ is already in row echelon form, $P$
and $U$ are identities. The pattern is isomorphic to the iteration space,
and $B$ is the fitting matrix.

\section{The Parametric Case} \label{approx}

Parameters occurs mostly in loop bounds. They may also appear as strides
and, more seldom, in the coefficients of subscript functions.

In the Array-OL formalism, the repetition loops must be square. Hence, 
their bound may be extracted diretcly from the program text. The extraction
of the paving matrix is a simple derivative computation, which is an easy task
for a competent computer algebra system.

Similarly, the $B^k$ matrices are the result of a derivation, and
may contain parameters.

There are no restrictions on the inner loops. For the construction of the
pattern, one needs to know the vertices of the inner iteration domain.
There are three cases:
\begin{itemize}
 \item The bounds are constant: they can be extracted even if parametric.
 \item The bounds are affine expressions in other loop counters and 
parameters: the vertices can be computed with the help of the polylib.
 \item In other cases, there is no way of computing vertices, but the
user may supply a bounding box.
\end{itemize}

The computation of the row echelon form can be done only if the matrix
is known numerically, except in two cases: the matrix is 
$1 \times 1$ (it is its own normal form) or $2 \times 2$.

\begin{quote} \small
The row echelon form of
$\left( \begin{array}{cc} a & b \\ c & d \end{array} \right)$
is
$\left( \begin{array}{cc} \gcd(a,b) & 0 \\
                        cu + dv  & |(ad-bc)| / \gcd(a,b) \end{array} \right)$
where $u$ et $v$ are the integers such that $a u + b v = \gcd(a,b)$
whose existence is guaranteed by Bezout identity.
\end{quote}

If none of these circumstance applies, the solution of last resort is to
use one of the approximate schemes above. For instance, if the vertices
of the inner iteration domain are available, it is possible,
whatever the $B$ matrix, to compute
the vertices of the footprints and to enclose them in a rectangular box.
The paving matrix is then the identity.

\section{Extensions}

The Syntol tool computes dependences; it is thus possible to check that
the repetition loops are actually parallel. One must take care that
Syntol will find dependences if temporary scalars are used in the code
of the elementary transforms. These scalars must be expanded or
privatized at code generation time.

Overlap between patterns (or, rather, between footprints) is another
concern. For input arrays, overlap is just a cause of inefficiency, since
some arrays cells will be copied several times to processors. Overlap
for output arrays are more dangerous since they may induce non-determinism.
The existence of overlap may be tested provided one stays inside the
polytope model (affine loop bounds and indexing functions, with
numerical coefficients and linear parameters). In the same context, it is
possible to quantify the overhead by comparing the size of the pattern
and the size of the real footprint using the {\tt barvinok} library
\cite{Segh:06}.

\appendix

\section{Computing the row echelon form of a matrix}

For more details, see \cite{Schr:86}.
Let $B$ be an arbitrary matrix of size $p \times q$.
\begin{enumerate}
 \item At any stage of the computation, we have constructed two unitary
matrices $P$ and $U$ such that:
\[ B = P B' U, \; B' = \quatrice{H}{0}{C}{D}\]
where $H$ is lower triangular with positive diagonal coefficients. Initially,
$P$ and $U$ are identity matrices, $H$ and $C$ are empty and $D=B$.
Let $i$ be the index of the first row of $C$ and $D$.
 \item If $D$ is null, the process stops. \label{fin}
 \item If not, let $j$ be the index of some non zero row of $D$. Let $\pi_{ij}$
be the unitary matrix that permutes rows $i$ and $j$ of $B'$. Since 
$\pi_{ij}$ is its own inverse, one can write:
\[ B = (P \pi_{ij}) (\pi_{ij} B') U ,\]
and the new $D$ has a non zero first row.
 \item Let $k$ be the index of a negative element in the first row of $D$.
Let $\sigma_k$ be the unit matrix with the $k$-th diagonal element set to
$-1$. Since  $\sigma_k$ is its own inverse, one can write:
\[ B = P (B' \sigma_k) (\sigma_k U) ,\]
and element $k$ in the first row of $D$ is now positive.
 \item If all elements in the first row of $D$ are positive, let
$l$ be the index of the smallest element, and let $\pi_{il}$ be the
matrix that interchange columns $i$ and $l$ of $B'$. Again:
\[ B = P (B' \pi_{il}) (\pi_{il} U) \]
and now the first element of the first row of $D$ is smallest.
 \item Let $m>i$ be the index of some nonzero element in the first row of $D$.
Set $\alpha = B'_{im}\div B'_{ii}$. By construction, $\alpha > 0$. Let
$\kappa_{im}(\alpha)$ be the identity matrix with $-\alpha$ added in position
$(i,m)$. It is easy to see that the inverse of $\kappa_{im}(\alpha)$ is
$\kappa_{im}(-\alpha)$. Hence:
\[ B = P (B' \kappa_{im}(\alpha)) (\kappa_{im}(-\alpha) U) \]
and element $B'_{im}$ has been replaced by $B'_{im} \bmod B'_{ii}$.
 \item If the only non-zero element of the first row of $D$ is the first 
element, then $i$ can be increased by 1.
\end{enumerate}
These transformations must be applied until no further progress is possible
(i.e. when in case \ref{fin}). Matrix $B'$ is in the required form, and since
all the elementary matrices $\pi, \sigma$ and $\kappa$ are unitary, the
resulting $P$ and $U$ are unitary. In fact, $P$ is even a permutation
matrix.

\bibliography{local}

\end{document}